# Role of electron and hole doping in NdNi$_{1-x}$V$_x$O$_3$ Nanostructure

Raktima Basu[1,a], Reshma Kumawat[1], Mrinmay Sahu[1], Abu Bakkar Miah[2], Partha Mitra[2], and Goutam Dev Mukherjee[1]

[1]*National Centre for High Pressure Studies, Department of Physical Sciences, Indian Institute of Science Education and Research Kolkata, Mohanpur Campus, Mohanpur 741246, Nadia, West Bengal, India*

[2]*Department of Physical Sciences, Indian Institute of Science Education and Research Kolkata, Mohanpur Campus, Mohanpur 741246, Nadia, West Bengal, India*

*Abstract*

Neodymium nickelate, NdNiO$_3$ attracts attraction due to the simultaneous occurrence of several phase transitions around the same temperature. The electronic properties of NdNiO$_3$ are extremely complex as structural distortion, electron correlation, charge ordering, and orbital overlapping play significant roles in the transitions. We report the effects of electron and hole injection via doping a single 3*d* metal, V, in the NdNiO$_3$ nanostructures to understand the variations in the electronic properties without any structural distortion. A reversible resistivity modulation more than five orders of magnitude via hole doping and complete suppression of metal to insulator transition via electron doping is observed along with the switching of major charge carriers. The modulation of electronic properties without any structural distortion and external strain opens up new directions to consider the NdNi$_{1-x}$V$_x$O$_3$ nanostructures applicable as emerging electronic devices.

[a]Authors to whom correspondence should be addressed. Electronic addresses: raktimabasu14@gmail.com

## I. INTRODUCTION

Correlated transition metal oxides find tremendous attention in the scientific community due to their metal-insulator transition (MIT) as a function of various external stimuli like temperature, radiation, pressure, chemical doping, magnetic field, and others.[1] MIT, in the rare-earth nickelates ($RNiO_3$), attracted significant attention in recent years due to the complexity in the mechanism of phase transition.[2] $RNiO_3$ (R = Nd, La, Pr, Sm, Eu, Y, etc.) undergoes a first-order structural transition from an orthorhombic (*Pbnm*) to a monoclinic (*P2₁/n*) phase along with the MIT from metallic (without charge ordering) to insulating (rocksalt-type charge ordering) phase.[3] Additionally, a magnetic transition from paramagnetic to E'-antiferromagnetic phase also occurs at a lower temperature. Among the nickelates, $NdNiO_3$ (NNO) is particularly interesting because the magnetic transition and MIT occur at the same temperature (~200K).[4] Moreover, as both the transitions appear simultaneously in bulk NNO, the origin of the phase transition is an open question for the researchers. Structural transition (electron-phonon coupling), electron correlations (Mott-Hubbard/Charge-transfer), charge-ordering, and magnetic ordering are proposed to play a significant role in the transition.[5] Due to its metal to insulator transition and charge ordering phenomena, $NdNiO_3$ finds many technological applications like in thermal sensors, electrical switches, thermochromics coating, photovoltaic cells etc.[6-8]

To increase the $T_{MIT}$ near room temperature several studies have been carried out, especially, by doping at Nd site.[9] It is well known for the $RNiO_3$ systems that $T_{MIT}$ increases with decrease in R cation size because of the change in Ni–O–Ni bond angle and Ni-O bond lengths.[10,11] There are also reports of modifying the $T_{MIT}$ by reducing the dimension from 3D to 2D[12] and by inducing strain either via thickness variation[13] or via substrate variation.[14] However, there are only a few reports of doping other 3*d* transition metals at Ni site[15-18] in bulk and 2D films. Previous reports suggest that doping at Ni site leads to a distortion in $NiO_6$ octahedral unit by affecting the Ni-O bond length and Ni-O-Ni bond angle[15] due to the size mismatch between Ni and the doping metal. Doping may also alter the hybridization between $O_{2p}$ and $Ni_{3d}$ orbitals and the bandwidth of the system and hence the electronic properties.[15] $RNiO_3$ are considered as charge transfer insulators according to Zaanen-Sawatzky-Allen (ZSA) classification, where Coulomb repulsion energy U (gap between occupied and unoccupied $Ni_{3d}$ bands) is more than the charge transfer energy $\Delta$ (gap between $O_{2p}$ and $Ni_{3d}$ band)[15] Doping with Cu, Co and Fe at Ni sites follows Torrance classification[15] at higher concentrations, and hence Fe or Co substitution increases the insulating behaviour, and Cu substitution increases the metallic nature of the system.[16,17] However, at low concentrations, their behaviour is explained by occupational disorder induced by substitution and localization of the spin density wave.[16] Moreover, doping can be either electron-type or hole-type depending on the valency of the doping metal ion. Substitution of Zn and Sr at Ni site causes injection of holes and is reported to bring metallicity in the system.[18, 19] Whereas, doping with Mn suggests electron doping and leads to a complete insulating phase.[20] As, various factors including electron/hole doping, charge ordering, and octahedral distortion due to size/strain mismatch, comes into picture, it hence become complicated to find a unique reason to explain the role of doping on the electronic properties of the NNO system. Vanadium has valency ranging from +2 to +5, and therefore may induce either electron or hole depending on the concentration. Moreover, the previous engineering in NNO have been limited to either bulk or 2D epitaxial thin films, where strain distribution may be complicated due to misfit dislocations. On the other hand, single-crystal 1D nanostructures exhibit novel physical properties due to their uniaxial nature and are expected to be free from any significant strain arising from outsources. In the present study, we report the effect of hole and electron injection via doping a single 3$d$ metal, V, in the NNO nanostructures and observed switching of majority charge carriers and variations in the electronic properties without any structural distortion.

## II. EXPERIMENTAL DETAILS

$NdNi_{1-x}V_xO_3$ (x=0.00, 0.01, 0.02) samples were synthesized using solid-state method. $Nd_2O_3$ (Alfa Aesar, 99.9%), and NiO (Alfa Aesar, 99.9%) powders were mixed in stoichiometric ratios. The mixture was then grinded for 1h and then dissolved in concentrated $HNO_3$ for decomposition. The excess $HNO_3$ was removed by heating the solution at 100°C until it turns into green coloured gel. The V foil with desired doping percentage (named as $S_1$, $S_2$) was kept along with the gel in Alumina boat and slowly heated to 400°C for 1h (green coloured gel turns in to black powder) in the furnace. After cooling, the sample was taken out from the furnace and grinded well to make pellets at 5 Kbar. Finally, the pellets were heated at 1100°C for 3 days in an open furnace. Sample $S_0$ was grown without any doping. The morphology of the samples were studied using field emission scanning electron microscopy (FESEM, SUPRA 55 ZEISS) and the information about the elemental compositions of the samples were found using energy diffraction X-ray (EDX). The crystallographic analysis for the prepared samples were obtained using the Rigaku powder X-Ray diffractometer (PXRD) using Cu K$\alpha$1 emission line with wavelength $\lambda$ =1.5406 ˚A. The observed XRD patterns were analysed by Rietveld refinement method using GSAS program. The vibrational analysis was performed using a Micro-Raman spectrometer (Monovista from SI GmbH) in backscattering geometry with excitation wavelength of 532 nm (Coboltsamba diode pump laser). The scattered light is dispersed using a grating with 1500 grooves/mm having a spectral resolution of 1.2 cm$^{-1}$ and is collected by a CCD detector (back-illuminated PIXIS 100BR). The resistivity measurement as a function of temperature was carried out using the four-probe Van Der Pauw method. In this method, the four probes were placed at equal distance on the perimeter of the pallets of homogeneous thickness unlike the linear four-probe method. We performed the temperature dependent measurements using a liquid helium cryostat under high vacuum. Hall measurements were carried out at room temperature using a custom built setup consisting of a four-pole electromagnet (Dexing Magnet Tech. Co. Limited), where the two perpendicular components of magnetic field can be controlled independently using two programmable power supplies. Magnetic field is varied perpendicular to the sample up to a maximum of +/-0.56 T and the corresponding Hall voltage is measured using Van Der Pauw configuration. A DC current of 5mA is applied using Keithley 6221 AC-DC current source and voltage is measured using Keithley 2182A nano-voltmeter. The Hall resistance is extracted as the component anti-symmetric with respect to applied magnetic field.

## III. RESULTS AND DISCUSSION

The FESEM images of the pristine samples are shown in figure 1. All the three samples $S_0$, $S_1$ and $S_2$ show similar morphology with average size of nano-crystals ~200 nm (Fig 1).

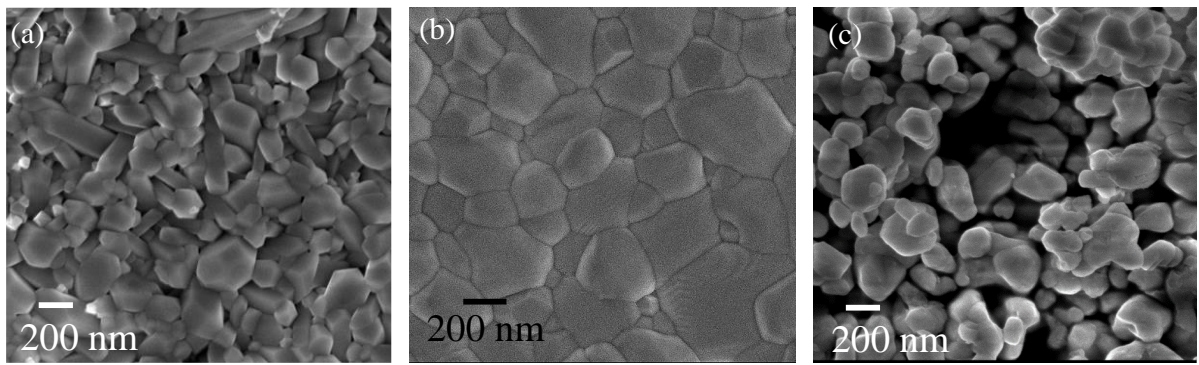

Figure 1. FESEM images for the pristine samples $S_0$, $S_1$ and $S_2$

The EDX spectra (figure 2) confirms the elemental composition of the samples. Corresponding atomic percentages of each element present in the sample are shown in the insets of fig. 2. In sample $S_0$ (figure 2a), Nd, Ni and O are in ratio 1:1:3 confirming the formation of phase pure $NdNiO_3$. However, for sample $S_1$ and $S_2$, the presence of V is observed, which mostly replaces Ni ion as $NdNi_{1-x}V_xO_3$ (x ~0.01 for $S_1$ and ~0.02 for $S_2$). The FESEM images (supplementary figure S1) and EDX spectra (supplementary figure S2) for higher doping percentages are shown in supplementary.

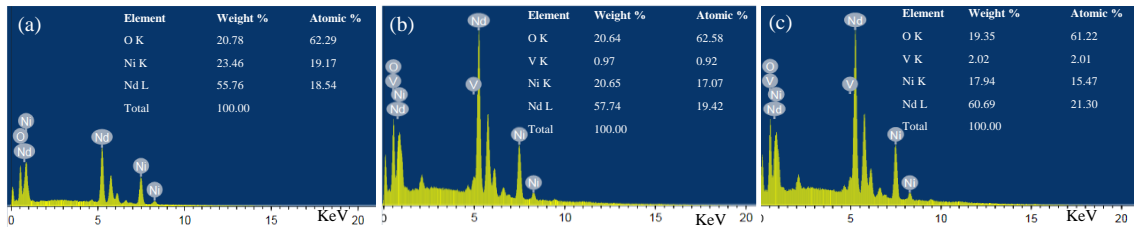

Figure 2. EDX spectra for the samples (a) $S_0$, (b) $S_1$ and (c) $S_2$. Insets show the elemental composition of each sample with corresponding atomic percentages.

The structural phase of the as-grown samples was analysed by PXRD using Bragg-Brentano geometry. Figure 3 shows the Rietveld refined PXRD pattern confirming orthorhombic $NdNi_{1-x}V_xO_3$ (x = 0, 0.01 and 0.02) phase (ICSD #111149) for the samples $S_0$, $S_1$, and $S_2$. However, for higher doping concentrations we observe the presence of $VO_x$ impurities along with NNO (supplementary figure S3).

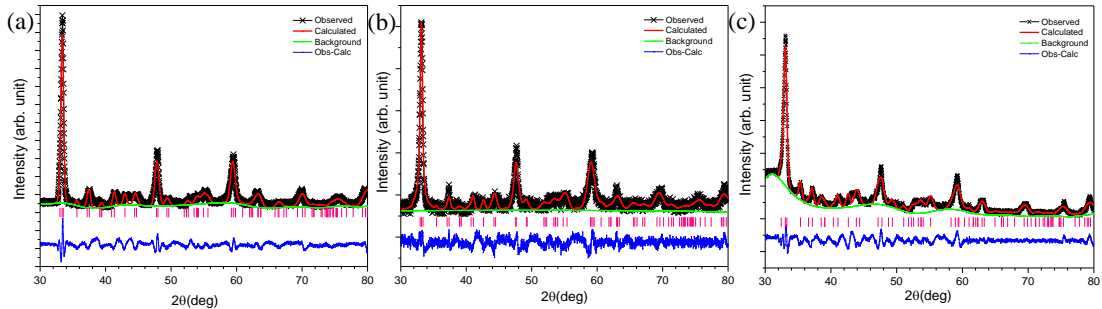

Figure 3. Rietveld refined X-ray diffraction pattern of samples (a) $S_0$, (b) $S_1$, (c) $S_2$.

From the refined patterns, we have found that the doped samples are fitted quite well with the phase pure NNO with goodness of fit (GOF, $\chi^2$) less than 5. The refined schematic unit cell structure of the samples were plotted using VESTA software (figure S4). The obtained lattice parameters, unit-cell volume, average bond-lengths, and bond-angles, along with fitting factors are tabulated in table 1. From the structure, we have observed minute changes in the average Ni-O bond lengths and Ni-O-Ni bond angles with doping concentrations (Table 1). The role of structural distortion in changing the electronic properties thus can be eliminated in this case. However, a reduction in unit cell volume is observed for the sample $S_1$, while a slight increase in volume is observed for sample $S_2$ as compared to the undoped sample $S_0$. As, V can be doped as $V^{+2}$ (hole), or $V^{+4}$ (electron), to maintain the total charge neutrality, $Ni^{+3}$ (ionic radius; r = 0.56 Å) also changes its valence state to smaller size $Ni^{+4}$ (r = 0.48 Å) or larger size $Ni^{+2}$ (r = 0.69 Å), accordingly.[20] The above observation suggests that sample $S_1$ may undergo divalent doping, while sample $S_2$ experiences tetravalent doping by V.[18]

Table 1: Refined Ni-O average bond lengths and Ni-O-Ni bond angles of NdNi$_{1-x}$V$_x$O$_3$ nanostructures

| Doping percentage (x) | Crystal system | Space group | Lattice parameters | | | R-factors | | Unit Cell Volume (Å$^3$) | Average Ni-O bond length (Å) | Ni-O-Ni bond angle (deg.) |
|---|---|---|---|---|---|---|---|---|---|---|
| | | | $a$(Å) | $b$(Å) | $c$(Å) | $R_p$(%) | $R_{wp}$(%) | | | |
| 0.00 | Orthorhombic | Pbnm | 5.4050 | 5.3689 | 7.6004 | 5.06 | 6.02 | 220.56 | 1.9441 | 154.71 |
| 0.01 | Orthorhombic | Pbnm | 5.3762 | 5.3767 | 7.5992 | 5.37 | 7.01 | 219.66 | 1.9434 | 154.23 |
| 0.02 | Orthorhombic | Pbnm | 5.3840 | 5.3840 | 7.6150 | 5.72 | 7.71 | 220.74 | 1.9642 | 154.54 |

At room temperature NdNiO$_3$ stabilizes in orthorhombic structure with space group $P$bnm. Group theory predicts 24 Raman active modes at Γ point; $7A_g + 7B_{1g} + 5B_{2g} + 5B_{3g}$. However, we observed only seven Raman modes due to the metallic nature of the sample at room temperature. The Raman spectra for the nanostructured NNO are reported for the first time, which resembles with its 2D and 3D counterparts except from the broad nature of the mode frequencies. The Raman mode frequencies at 205, 253, 303 404, 452, 462 cm$^{-1}$ and a broad peak after 500 cm$^{-1}$, match with the previously reported data and confirm the formation of NNO.[21,22] However, we observe an increase in Raman background for sample S$_1$, whereas, the background reduces for sample S$_2$ with respect to the un-doped sample S$_0$. An increase (decrease) in Raman background indicates increase (decrease) in metallicity in the system.[23] Hole doping in NNO is reported to bring metallicity in the system, whereas electron doping leads to decrease in metallicity.[18,20,22-24] Our observations from Raman spectroscopic studies indicate that sample S$_1$ is doped by hole (V$^{+2}$), while in sample S$_2$ the substitution is via electron (V$^{+4}$), as also presumed from XRD refinement studies (table 1).

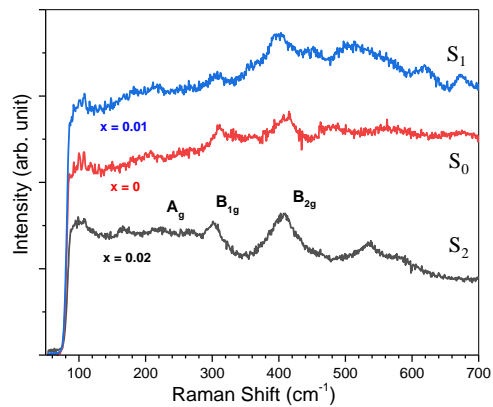

Figure 4. Raman spectra of NdNi$_{1-x}$V$_x$O$_3$ nanostructures with variation in doping concentrations.

To study the phase transition, we have carried out the resistivity measurements as a function of temperature. Figure 5 shows the resistivity curves for NdNi$_{1-x}$V$_x$O$_3$ nanostructures with temperature from 50 to 300K. For sample S$_0$, there is a discontinuous fall in resistivity up to three orders in magnitude at 110K, which matches with the previous reports for that of NNO thin films.[25] Sample S$_1$ also shows the drastic fall in resistance around the same temperature. However, for sample $S_1$, the resistivity curve shows a sharp drop of five orders in magnitude unlike thermally driven ~ one-two orders in bulk NdNiO$_3$.[3,26] The resistivity value at room temperature for sample $S_1$ is two orders lower than that of undoped NNO ($S_0$), indicating that sample $S_1$ is more metallic than sample $S_0$. However, for sample $S_2$, we observe a continuous fall in resistivity with increasing temperature suggestive of insulating behaviour. The resistivity value at room temperature is two orders higher than sample $S_0$, indicating a lower metallicity for sample $S_2$. The resistivity study also shows that the transition is reversible with a hysteresis about 10K (supplementary figure S5). Moreover, for sample S1, we observe another transition at ~225K, leading to the emergence of a new metallic phase. $V^{2+}$ doping in sample $S_1$ inserts more carriers in the system, and thereby broadens the electronic bands. The broadening of bands leads to overlapping of oxygen 2$p$ and Ni 3$d$ orbitals making the system more metallic.

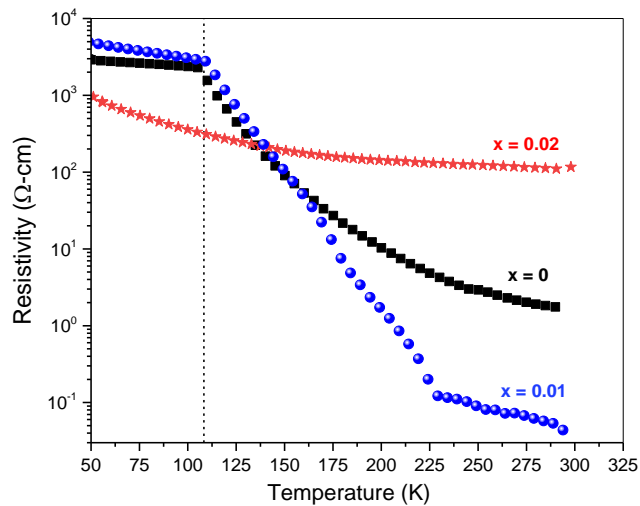

Figure 5. Resistivity plots as a function of temperature for NdNi$_{1-x}$V$_x$O$_3$ (x = 0.00, 0.01 and 0.01) nanostructure. The vertical line indicates the transition temperature.

The nature of carrier in rare-earth nickelates may be either holes in the oxygen $p$ band, or electrons in the transition metal $d$ band.[27,28] To find the carrier density in the samples we have performed Hall resistivity measurements for un-doped and doped samples. However, the data looks more scattered for the doped samples. Possibly higher magnetic field is needed to achieve a better symmetric behaviour. Our experimental set up is limited by a maximum magnetic field of 0.56 T. Therefore, we could not carry out the Hall measurement for magnetic field beyond 0.56 T. However, a general trend could be seen by our attempt using a linear fit (details are given in supplementary figure S6). The result shows that the major charge carrier in sample $S_0$ is hole and the carrier density is ~1.54 x 10$^{27}$ m$^{-3}$. Sample $S_1$ shows carrier density ~2.18 x 10$^{27}$ m$^{-3}$. The increase in carrier density indicates that $V^{2+}$ doping injects holes in the 3$d$ band and therefore, there is an enhancement in the density of charge carriers in the system. However, for sample $S_2$, the carrier density ($3.31 \times 10^{26}$ m$^{-3}$) decreases by one order along with an electron-like doping in the system. The change in the switching of majority charge carriers from hole-like to electron-like in sample $S_2$ makes the sample insulating in nature. Temperature-induced switching of charge carriers has already been reported for NdNiO$_3$ system.[29,30] However, doping induced switching using a single 3*d* metal eliminates the structural-distortion induced charge transfer, while an electron/hole doping induced band overlapping is found to be more prominent in controlling the electronic properties in NdNi$_{1-x}$Zn$_x$O$_3$ nanostructures.

## IV. CONCLUSION:

In summary, the structural and electronic properties of NdNi$_{1-x}$V$_x$O$_3$ (x = 0.00, 0.01, 0.02) nanostructures have been carried out. We report the effect of electron and hole injection via doping a single 3*d* metal, V, in the NNO system to understand the phase transition with major charge carrier switching and variations in the electronic properties. Hole doping leads to over five orders magnitude drop in resistivity and emergence of a new metallic phase. Whereas, electron doping leads to suppression of the phase transition and induces insulator-like nature in the system. Our study reveals that engineering of phase transition induced by structural-distortion can be eliminated by using a single dopant for both electron and hole doping. Doping induced modulation of electronic properties without any structural distortion and external strain makes the NdNi$_{1-x}$V$_x$O$_3$ nanostructures applicable as emerging electronic/ionic devices.

## SUPPLEMENTARY MATERIAL

See the supplementary material for additional data including FESEM, EDX, PXRD, Schematic structures from Rietveld refined XRD patterns, Resistivity and Hall measurements.

## ACKNOWLEDGEMENTS

We acknowledge the financial support from SERB, DST, Government of India. We also thank Ms. Piyali Ghosh for collecting XRD data and Mr. Kashi Nath Sahu for performing FESEM and EDX measurements.

## AUTHOR DECLARATIONS

**Conflict of Interest**

All the authors declare no competing financial interest.

## DATA AVAILABILITY

The data that support the findings of this study are available from the corresponding author upon reasonable request.